\documentclass[pre,floatfix,a4paper,nofootinbib,amssymb,amsmath,showpacs,superscriptaddress]{revtex4}
\usepackage{graphicx}

\newcommand {\be} {\begin{equation}}
\newcommand {\ee} {\end{equation}}
\newcommand {\Be}{\begin{eqnarray*}}
\newcommand {\Ee} {\end{eqnarray*}}
\newcommand {\bey} {\begin{eqnarray}}
\newcommand {\eey} {\end{eqnarray}}
\newcommand{\bit}{\begin{itemize}}      
\newcommand{\eit}{\end{itemize}}
\newcommand{\bfl}{\begin{flusleft}}
\newcommand{\efl}{\end{flusleft}}
\newcommand{\bfr}{\begin{flushright}}
\newcommand{\ec}{\end{center}}
\newcommand{\ben}{\begin{enumerate}}    
\newcommand{\een}{\end{enumerate}}

\newcommand{\comment}[1]{}

\begin{document} 

\title{Collective oscillations in disordered neural networks}

\author{Simona Olmi}
\email{simona.olmi@fi.isc.cnr.it}
\affiliation{Physics Department, via Sansone, 1 - I-50019 Sesto Fiorentino, Italy}
\affiliation{Istituto dei Sistemi Complessi, CNR, via Madonna del Piano 10, I-50019 Sesto Fiorentino, Italy}
\affiliation{INFN Sez. Firenze, via Sansone, 1 - I-50019 Sesto Fiorentino, Italy}
\affiliation{Centro Interdipartimentale per lo Studio delle Dinamiche Complesse, via Sansone, 1 - I-50019 Sesto Fiorentino, Italy}
\author{Roberto Livi}
\email{livi@fi.infn.it}
\affiliation{Physics Department, via Sansone, 1 - I-50019 Sesto Fiorentino, Italy}
\affiliation{Istituto dei Sistemi Complessi, CNR, via Madonna del Piano 10, I-50019 Sesto Fiorentino, Italy}
\affiliation{INFN Sez. Firenze, via Sansone, 1 - I-50019 Sesto Fiorentino, Italy}
\affiliation{Centro Interdipartimentale per lo Studio delle Dinamiche
Complesse, via Sansone, 1 - I-50019 Sesto Fiorentino, Italy}
\author{Antonio Politi}
\email{antonio.politi@cnr.it}
\affiliation{Istituto dei Sistemi Complessi, CNR, via Madonna del Piano 10, I-50019 Sesto Fiorentino, Italy}
\affiliation{Centro Interdipartimentale per lo Studio delle Dinamiche
Complesse, via Sansone, 1 - I-50019 Sesto Fiorentino, Italy}
\author{Alessandro Torcini}
\email{alessandro.torcini@cnr.it}
\affiliation{Istituto dei Sistemi Complessi, CNR, via Madonna del Piano 10, I-50019 Sesto Fiorentino, Italy}
\affiliation{INFN Sez. Firenze, via Sansone, 1 - I-50019 Sesto Fiorentino, Italy}
\affiliation{Centro Interdipartimentale per lo Studio delle Dinamiche
Complesse, via Sansone, 1 - I-50019 Sesto Fiorentino, Italy}

\begin{abstract}
We investigate the onset of collective oscillations in a network of
pulse-coupled leaky-integrate-and-fire neurons in the presence of 
quenched and annealed disorder. We find that the disorder induces
a weak form of chaos that is analogous to that arising in the Kuramoto model
for a finite number $N$ of oscillators [{\it O.V. Popovich at al., Phys. Rev. E {\bf 71} 065201(R) (2005)}].
In fact, the maximum Lyapunov exponent turns out to scale to zero for $N\to\infty$, with an exponent that
is different for the two types of disorder. In the thermodynamic limit,
the random-network dynamics reduces to that of a fully homogenous
system with a suitably scaled coupling strength. Moreover, we
show that the Lyapunov spectrum of the periodically collective state
scales to zero as $1/N^2$, analogously to the scaling found for the
`splay state'.

\end{abstract}
   

\maketitle


\section{Introduction}\label{one}

One of the most general and relevant dynamical phenomena observed in the
mammalian brain is the rythmic coherent behaviour involving different neuronal
populations \cite{buszaki}. The dynamics of neural circuits has been widely
studied, by invoking various kinds of neuron models; collective oscillations
are commonly associated with the inhibitory role of interneurons
\cite{interneurons}. However, coherent activity patterns have been observed also
in ``in vivo'' measurements of the developing rodent neocortex and hyppocampus
for a short period after birth \cite{allene}, despite the fact that at this early
stage the nature of the involved synapses is essentially excitatory \cite{review}.

Independently, theoretical studies of fully coupled excitatory networks of
leaky integrate-and-fire (LIF) neurons have revealed the onset of macroscopic
collective periodic oscillations~\cite{vvres,mohanty} (CPOs). This dynamical state
is quite peculiar: the collective oscillations are a manifestation of a
{\it partial synchronization} among the neuron dynamics and this is one way of
identifying this phenomenon, which is, however, more subtle: the macroscopic
period of the oscillations does not coincide with (is longer than) the average
interspike-interval ISI of the single neurons and the two quantities are
irrationally related. In fact, this phenomenon is also called self-organized
quasi periodicity and can be observed in a wide class of globally coupled
systems \cite{piko}. In the context of
pulse-coupled neural networks, CPOs arise from the destabilization of a 
regime characterized by a constant mean-field and a strictly periodic evolution
of the single neurons: this regime, termed  ``splay state'',  has been widely
studied in several contexts, including computational neuroscience \cite{abbott}.

Since real neural circuits are not expected to have a full connectivity
\cite{koch}, it is important to investigate the role of dilution on the
occurrence of the stability of CPO. We do so by investigating an excitatory
network of LIF neurons with 20\% of missing links in two different setups:
{\it quenched} disorder, where the network topology is fixed; {\it annealed}
disorder, where the active connections are randomly and independently chosen at
each pulse emission. As a first step, we rewrite the dynamical
equations as a suitable event driven map, by extending the approach developed
in \cite{zillmer2}. We do so by introducing a pair of variables for each neuron,
to account for the evolution of the local electric field. This step is
particularly important for the computation of the Lyapunov exponents, as it
allows expressing the evolution equations into a ``canonical" form and thereby
simplifies the implementation of standard dynamical-system tools.

We find that the regime of CPOs is robust against the presence of dilution,
both in the quenched and annealed setup. However, at variance with the
homogeneous fully-coupled case, the dynamics of finite disordered networks
turns out to be chaotic, although the degree of chaoticity decreases with the
number $N$ of neurons. In fact, the maximum Lyapunov exponent goes to zero as
$1/N^\beta$. The exponent $\beta$ is smaller in the quenched setup, indicating
that finite-size effects are stronger. In the homogeneous case, we are able
to determine the full Lyapunov spectrum for sufficiently large numbers of
neurons. As a result, we find that the first band of the spectrum 
scales as that of the splay state \cite{abbott,calamai}, namely as $1/N^2$

The paper is organized as follows. In section \ref{two} we introduce the model
and the event-driven map that is used to carry out the stability analysis.
In Sec.~\ref{three} we discuss the collective dynamical behaviours observed in
the presence of disorder. Sec.~\ref{four} is devoted to the Lyapunov analysis of
these coherent solutions in the large $N$ limit. Finally, in Sec.~\ref{five}, we
summarize the main results and the open problems.


\section{The model}\label{two}

We study a network of $N$ neurons, whose individual dynamics is modelled as a
LIF oscillator. Following Refs.~\cite{zillmer2}, the membrane potential $x_i(t)$
of the $i-th$ oscillator evolves according to the differential equation
\begin{equation}\label{eq:x1}
  \dot{x}_{i}(t)= a-x_{i}(t)+g \, E_i(t)\, \quad\quad i=1,\cdots, N
\\
\end{equation}
where $a >1$ is the suprathreshold input current and $g > 0$ gauges the 
coupling strength of the excitatory interaction with the neural field 
$E_i(t)$. At variance with the fully-coupled network, where all
neurons depend on the same ``mean field" $E(t)$, here we consider a general
setup, where neurons have different connectivities inside the network. 
As a result, it is necessary and sufficient to introduce an explicit dependence
of the neural field on the index $i$. The discharge mechanism operating in real
neurons is modelled by assuming that when the membrane potential
reaches the threshold value $x_i=1$, it is reset to the value $x_i=0$,
while a pulse is transmitted to and instantaneously received by the connected
neurons. The field $E_i(t)$ is represented as the linear superposition of the
pulses received by neuron $i$ at all times $t_n < t$: the integer index $n$ 
orders the sequence of the received pulses. Each pulse is weighted according to
the strength of the connection $C_{j,i}$ between the emitting ($j(n)$) and the
receiving ($i$) neuron. In general, the connectivity matrix $\bf C$ is assumed
to be non-symmetric. In formulae,
\begin{equation}\label{eq:E0}
 E_i(t)= \frac{\alpha^2}{N} \sum_{n|t_n<t} C_{j(n),i} (t-t_n)
\theta\bigg(t-t_n^{(i)}\bigg) \exp \left[-\alpha (t-t_n)\right]
\end{equation}
where $\theta(x)$ is the Heavyside function and $\alpha$ is the
inverse pulsewidth (the pulse shape has been chosen as in
Ref.~\cite{vvres,abbott})~. Since also in the diluted case
we will consider massively connected networks \cite{hansel}, 
where the average number of synaptic inputs per neuron varies proportionally to
the system size, it is natural to scale the synaptic strength with $N$ as
done in Eq. (\ref{eq:E0}).

The model is fully characterized by the two sets of equations (\ref{eq:x1}) and
(\ref{eq:E0}). The dynamical system has a peculiar mathematical structure, with
the field variable appearing as a memory kernel, which involves a summation
over all past spiking events. 
We find much more convenient to turn the explicit equation (\ref{eq:E0}) into
the implicit differential equation
\begin{equation}\label{eq:E}
  \ddot E_i(t) +2\alpha\dot E_i(t)+\alpha^2 E_i(t)= 
  \frac{\alpha^2}{N}\sum_{n|t_n<t} C_{j(n),i} \delta(t-t_n) \ .
\end{equation}

As a result, the dynamics of the neural network model takes the more
``canonical" form of a set of coupled ordinary differential equations
(\ref{eq:x1},\ref{eq:E}), which can be analyzed with the standard methods of
dynamical systems (see e.g., \cite{vvres,abbott,zillmer2,calamai})~.

\subsection{Event-driven map}

The presence of $\delta$-like pulses into the set of coupled differential
equations (\ref{eq:x1}) and (\ref{eq:E}) may still appear as an intrinsic
technical difficulty for the estimation of the stability properties.
Actually, the standard algorithms for the evaluation of Lyapunov exponents rely
upon the integration of differentiable operators acting in tangent space
(see \cite{benettin}). However, one can easily get rid of this problem by
tansforming the differential equations into a discrete time event-driven
mapping.
This task can be accomplished by integrating Eq.~(\ref{eq:E}) from time $t_n$ to 
time $t_{n+1}$
($t_n$ representing the time immediately after the emission of the $n$-th pulse).
An explicit mapping can be written by introducing the auxiliary variable 
$Q_i := \alpha E_i+\dot E_i$,
\begin{subequations}\label{eq:map}
\begin{gather}
  E_i(n+1)=E_i(n) {\rm e}^{-\alpha \tau(n)}+Q_i(n)\tau(n) 
  {\rm e}^{-\alpha \tau(n)} \\
  Q_i(n+1)=Q_i(n)e^{-\alpha \tau(n)}+C_{j(n+1),i}\frac{\alpha^2}{N} \\
  x_{i}(n+1)=x_i(n)e^{-\tau(n)}+a(1-e^{-\tau(n)})+g H_i(n) \, .
\end{gather}
\end{subequations}
Here $\tau(n)= t_{n+1}-t_n$ is the interspike time interval: it 
is determined by the largest membrane potential (identified by the label
$m(n)$) reaching the threshold value $x_m = 1$ at time $t_{n+1}$,
\begin{equation}\label{eq:ti2}
  \tau(n)=\ln\left[\frac{a-x_m(n)}{a+gH_m(n)-1}\right]\ ,
\end{equation}
where 
\begin{equation}\label{eq:F1}
  H_i(n)= \frac{{\rm e}^{-\tau(n)} - e^{-\alpha\tau(n)}}{\alpha-1}
     \left(E_i(n)+\frac{Q_i(n)}{\alpha-1} \right) - 
  \frac{\tau(n) e^{-\alpha\tau(n)}}{(\alpha-1)} Q_i(n) \, ;
\end{equation}
for the parameter values considered in this paper ($g > 0$ and $a \leq 1$),
$H_i(n) > 0$.

Altogether, the model now reads as a discrete time map of $3 N - 1$
variables, $\{E_i,Q_i,x_i\}$: one degree of freedom, $x_{m(n)} = 1$, is
eliminated as a result of having constructed the discrete-time dynamics 
with reference to a suitable Poincar\'e-section. At variance with the usual
approach (see e.g. Ref.~\cite{coombes}), the evolution equation does neither
involve $\delta$-like temporal discontinuities, nor formally infinite
sequences of past events: the map is a piecewise smooth dynamical system.  

For the sake of simplicity, we assume that the entries of the connectivity 
matrix $C_{j,i}$ are either 0 or 1 (the homogeneous fully-coupled case
correspondingto $C_{j,i} = 1$ for all $j$'s  and $i$'s). If the entries are
chosen randomly, symmetries are lost and the only way to determine the
asymptotic state for a finite $N$ is by numerically simulating map
(\ref{eq:map}). In what follows, we consider two different setups:
the synaptic connections are randomly chosen and are constant in time
(quenched disorder); each time a neuron fires, the neurons receiving the
excitatory pulse are randomly chosen (annealed disorder).

\subsection{Linear Stability Analysis}

As usual, the stability of (\ref{eq:map}) can be analyzed by 
following the evolution of infinitesimal perturbations in the tangent space.
The corresponding equations are obtained by linearizing (\ref{eq:map}) as
follows,
\begin{subequations}
\label{eq:lin1}
\begin{gather}
\delta E_i(n+1)=e^{-\alpha \tau(n)}\delta E_i(n)+ \tau(n) e^{-\alpha \tau(n)}\delta Q_i(n)
  -\left(\alpha E_i(n)(\alpha \tau(n)-1) Q_i(n)\right) e^{-\alpha \tau(n)}\delta \tau(n)\,,\\
  \delta Q_i(n+1)=e^{-\alpha \tau(n)} \left[ \delta Q_i(n)-\alpha Q_i(n) \right]
  \delta \tau(n)\, ,\\
  \delta x_{i}(n+1)=e^{-\tau(n)}[\delta x_i (n)- x_i(n) \delta \tau(n)]+ a e^{-\tau}
  \delta \tau(n) +g \delta H_i (n) 
\quad i=1,\dots,N-1 \quad ; \quad \delta x_m(n+1) \equiv 0 \, .
\end{gather}
\end{subequations}
An explicit expression of $\delta \tau(n)$ can be obtained by differentiating
Eqs.~(\ref{eq:ti2},\ref{eq:F1}) 
\begin{equation}
\label{eq:tauder}
  \delta \tau(n) =\tau_x \delta x_1(n) +\tau_E\delta E(n)+\tau_Q\delta Q(n)\ ,
\end{equation}
where $\tau_x:=\partial \tau/\partial x_1$ and analogous definitions are
adopted for $\tau_E$ and $\tau_Q$. Moreover, $\delta H_i (n)$ is 
a short-cut notation for the linearizazion of expression (\ref{eq:F1}),
which in turn depends on $\delta E_i(n), \delta Q_i(n)$ and $\delta \tau(n)$.
For more mathematical details see \cite{zillmer,zillmer2,calamai}.

The degree of chaoticity of a given dynamical state is obtained by
computing the
Lyapunov spectrum, i.e. the set of $3N -1$ exponential growth rates  
$\lambda_i$ along the independent directions in tangent space.
The Lyapunov spectrum has been numerically estimated by implementing the
standard algorithm \cite{benettin}.

\section{Collective Dynamics}
\label{three}

In the fully coupled homogeneous case the local fields $E_i$, $Q_i$ 
are independent of the index $i$ and the number of equations reduces to $N+1$. 
Depending whether $\alpha$ is smaller or larger than a critical
value $\alpha_c(g)$, the dynamics either converges to a so-called splay state,
with constant $E(t)$, or to a partially synchronized state, where
$E(t)$ and $x_i(t)$ evolve periodically and quasi-periodically, respectively
\cite{vvres,mohanty}. 

This ``mean-field" dynamics is expected to change in disordered networks. Given
the neuron dependence of the fields $E_i$ and $Q_i$, we find it convenient to 
introduce the average variables,
\begin{equation}
\label{eq:avefields}
  \bar E(n) = \frac{1}{N} \sum_{k=1}^N E_k(n) \qquad ; \qquad
  \bar Q(n) = \frac{1}{N} \sum_{k=1}^N Q_k(n)  \ .
\end{equation}
as a tool to characterize the resulting dynamical regimes and to compare with
the homogeneous case. All the results reported in this paper have been
obtained by fixing the fraction $f$ of missing connections equal to
$f = 0.2$. Numerical investigation for different choices of $f$ (not reported
here) show similar results.
In analogy with the analysis of the homogeneous case, we have chosen
$\alpha$ as the main control parameter (one can easily realize that choosing $g$
would be equivalent). 

\begin{figure}[t!]
\includegraphics[draft=false,clip=true,height=0.34\textwidth]{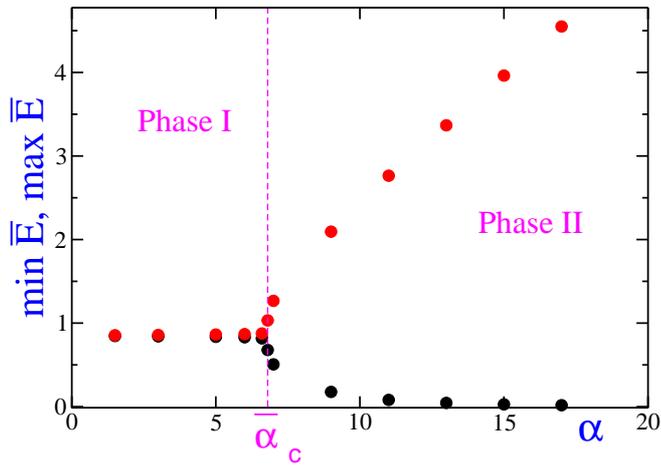}
\caption{Minimal and maximal values of $\bar E(n)$ for different
$\alpha$ values for $N=1,600$ with quenched disorder and $g=0.4$.}
\label{fig1}
\end{figure}

In Fig.~\ref{fig1} we plot the maximum and the minimum values of $\bar E(n)$ 
for different values of $\alpha$, for $g=0.4$ and $N = 1600$ in the presence of
quenched disorder. The bifurcation diagram is similar to the one
observed in the globally coupled networks \cite{vvres}. However, the
splay state found for small $\alpha$-values has been replaced by a fluctuating
asynchronous state, where the average field $\bar E(t)$ is only approximately
constant (the difference between  $\min \bar E(n)$ and $\max \bar E(n)$ is of
the size
of the symbols). The periodic collective state is analogously affected
by small irregular fluctuations. This strong similarity between 
globally coupled and the diluted networks is not surprising: for any finite
value of $f$, upon increasing $N$, the differences among the fields $E_i(t)$
should progressively disappear. In fact, in the limit $N\to\infty$, 
we expect randomly diluted networks to behave as fully coupled ones, provided
that variables and parameters are properly rescaled. More precisely, the
average field $\bar E(t)$ in a network with a fraction $f$ of missing links and
a coupling constant $g$, is expected to be equivalent to the neural field $E$
generated in a fully coupled network with coupling constant $g(1-f)$. This
expectation is confirmed by the critical value $\bar \alpha_c$ separating the
two dynamical phases. As shown in Fig.~\ref{fig1}, for $g=0.4$ and $f=0.2$,
$\bar \alpha_c \sim  6.8$, a value which coincides, within the statistical
error, with the critical value found in a globally coupled network with
$g=0.32 = 0.4\times (1-0.2)$\cite{zillmer2}.
A further consequence of the fact that the dynamics of disordered networks
reduces in the thermodynamic limit to that of homogenous fully coupled ones is
that the sample-to-sample fluctuations typical of quenched disorder should
vanish as well. For this reason we have not performed averages of different
realizations of the disorder.

\begin{figure}[t!]
\includegraphics[draft=false,clip=true,height=0.34\textwidth]{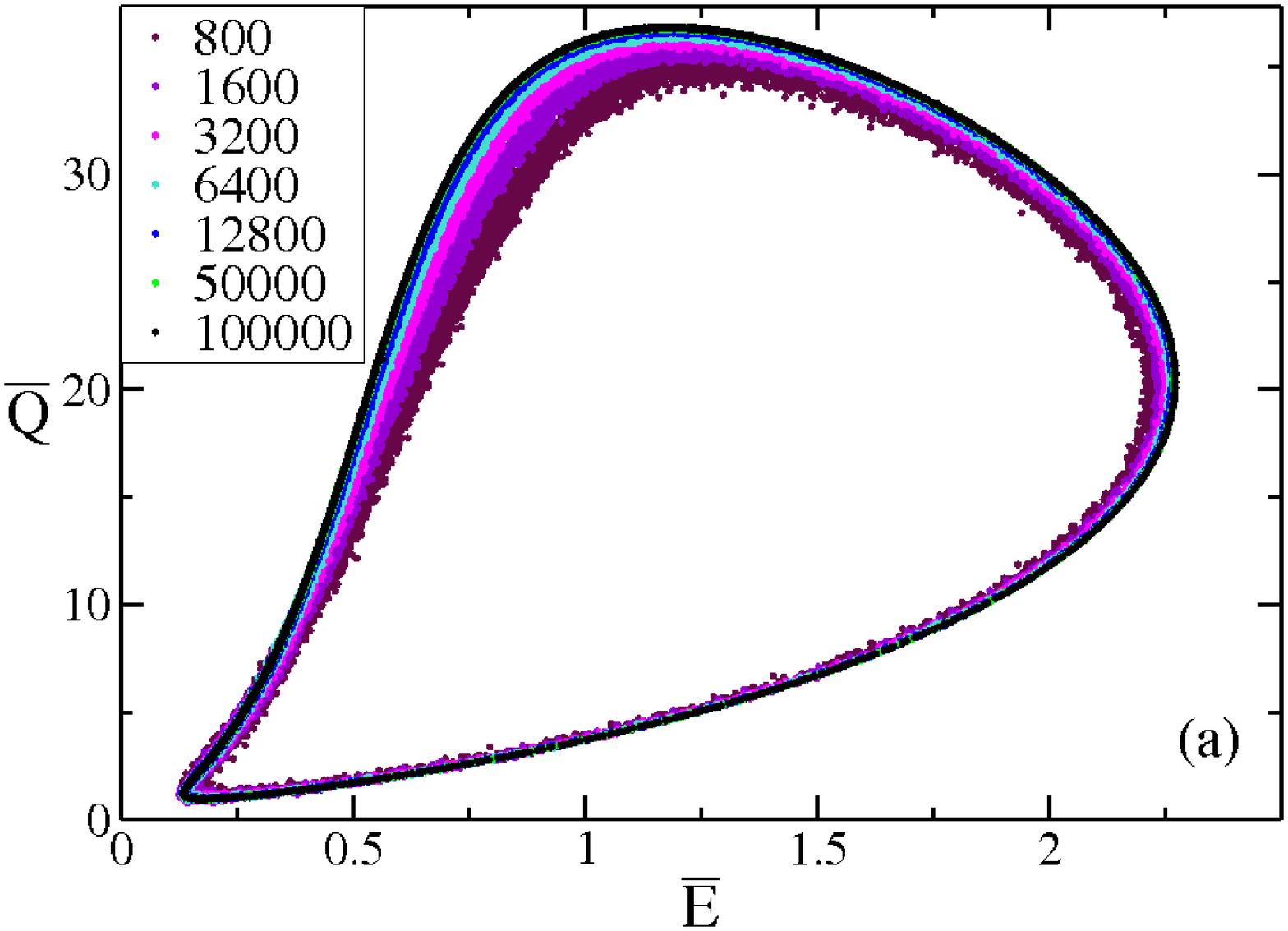}
\includegraphics[draft=false,clip=true,height=0.34\textwidth]{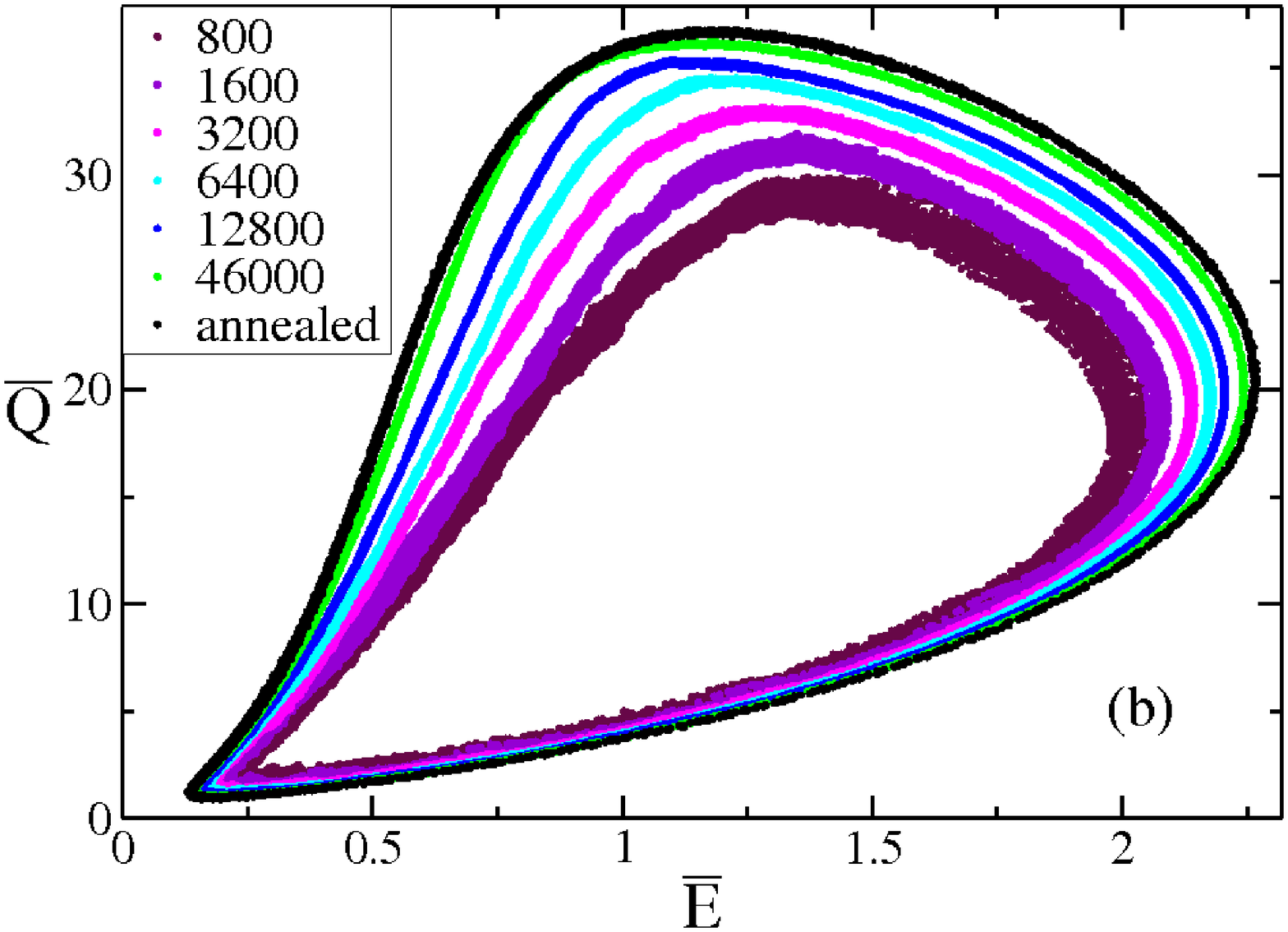}
\caption{(Color online) $\bar E(n)$  versus $\bar Q(n)$ for various system syze:
(a) annealed disorder and (b) quenched case here is shown also the annealed
result for $N=100,000$. All data refers to $\alpha=9$.
}
\label{fig2}
\end{figure}

A more detailed representation of the quenched dynamics above $\bar \alpha_c$
is obtained by looking at the projection in the plane $(\bar E(n), \bar Q(n))$.
In Fig.~\ref{fig2} data sets are shown for $\alpha=9$ and increasing values of
$N$: panels a and b correspond to the annealed and quenched case, respectively.
This allows seeing that the two kinds of disorder yield indeed qualitatively
similar results.

More precisely, the phase points cluster around closed curves, revealing a
``noisy" periodic dynamics. In the annealed case, the fluctuations can be
attributed to the stochasticity of the evolution rule. Surprizingly, they
are even larger in the quenched case, which correspond to a deterministic
dynamics, in which case they must be attributed to the presence of
deterministic chaos. 

\begin{figure}[ht]
\includegraphics[draft=false,clip=true,height=0.37\textwidth]{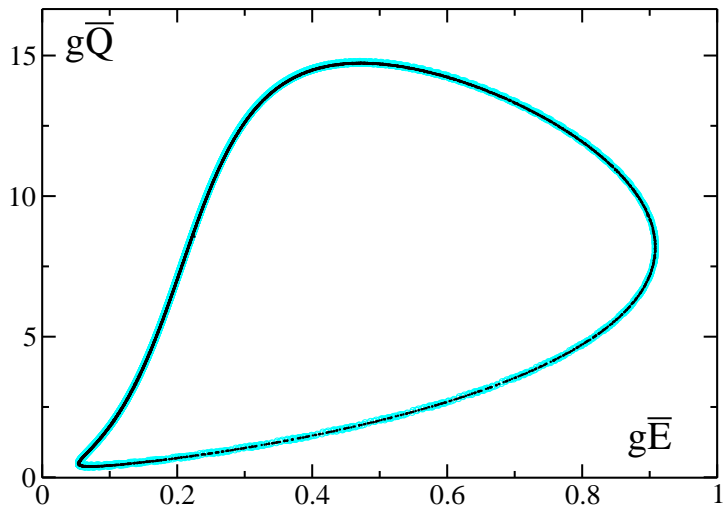}
\caption{ $g \bar E(n)$  versus $g \bar Q(n)$.
The (blue) curve represents the attractor of a globally coupled
system with $N=1,600$ and $g=0.32$, while the black dots refer to
the annealed case with $N=100,000$ 
and $g=0.4$. For both cases $\alpha= 9$ and $a=1.3$.}
\label{fig3}
\end{figure}

Upon increasing $N$, the amplitude of the fluctuations decreases and the
attractor shape converges to an asymptotic curve, which should correspond to
that of a homogeneous fully coupled network with a properly rescaled coupling
strength $g$. This expectation is indeed confirmed in Fig.~\ref{fig3}, where the
attractor of a homogeneous network (with $N=1600$ and $g=0.32$) superposes to
that of the annealed dynamics (with $N=10^5$ and $g=0.4$). In the quenched case,
the asymptotic shape is the same, but the convergence is slower.

\begin{figure}[ht]
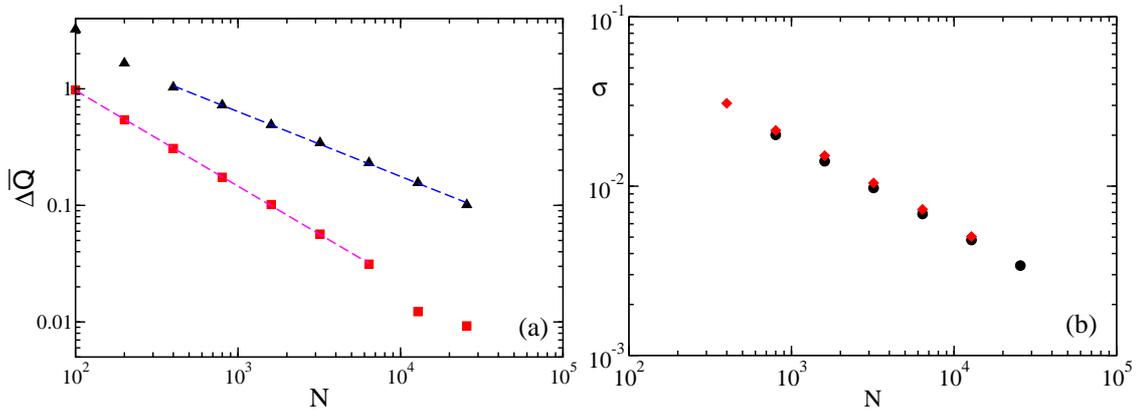

\includegraphics[draft=false,clip=true,height=0.29\textwidth]{FIG/fig4a}
\includegraphics[draft=false,clip=true,height=0.3\textwidth]{FIG/fig4b}
\caption{(Color online) (a) The quantity $\Delta \bar Q$ (see Eq. \ref{eq:deltaq}) versus the network size; 
(black) triangles refer to the quenched case, while (red) squares to the annealed one:
data fits
are reported as dashed lines. The reported values have been obtained 
by averaging over $6 \times 10^6$ to $3 \times 10^7$ consecutive spikes. (b)
Standard deviation $\sigma$ (see Eq.~\ref{eq:sigma}) vs $N$ : circles and
diamonds refer to the annealed and quenched case, respectively. The standard
deviation have been averaged over $M=900$ iterates. All data refers to $f=0.2$,
$\alpha= 9$, $g=0.4$ and $a=1.3$.
}
\label{fig4}
\end{figure}

In order to investigate quantitatively the scaling behavior of the finite-size
corrections, we have studied the $N$ dependence of 
\begin{equation}
\Delta {\bar Q} = \langle {\bar Q} \rangle(N) - \langle {\bar Q} \rangle(\infty)
\label{eq:deltaq}
\end{equation}
where the angular brackets denote the (time) average of $\bar Q$-value of all 
configurations falling within the lower $g \bar E(n)$-window
[0.36, 0.44]. Since the asymptotic value $\langle {\bar Q} \rangle(\infty)$ is
independent of the setup, we have extrapolated it in the simpler context of
a fully coupled network with $g=0.32$ As a result, we find that 
$\Delta {\bar Q}$ always converges to zero as  power-law, $N^{- \beta}$, with
$\beta=1$ in the fully coupled network, $\beta=0.55$ for quenched disorder,
and $\beta=0.82$ for annealed disorder (see Fig~\ref{fig4}a). 
These latter values have to be considered as approximate estimates,
affected both by statistical errors and finite-size corrections. A more
accurate estimate is beyond our computational capabilities. However, for the
present purpose, the relative differences are large enough to conclude that
quenched disorder is characterized by a slower convergence.

As a further test of the collective dynamics, we have computed the standard
deviation $\sigma(n)$ of the instantaneous fields
\begin{equation}
\label{eq:sigma}
\sigma(n) = \left( \frac{\sum_{i=1}^{N}E_{i}^{2}({n})}{N}-\bar{E}^{2}({n})\right)^{1/2} \, ,
\end{equation}
which is a measure of the degree of synchronization among the various fields.
In Fig.~\ref{fig4}b  we plot the time average $\langle \sigma \rangle$ for annealed
and quenched disorder: in both cases $\langle \sigma \rangle$ decreases with $N$
as $N^{-1/2}$. These results confirm once more that in the limit $N \to \infty$
the neural field dynamics converges to that of homogeneous networks,
irrespectively of the disorder.

\section{Lyapunov analysis}
\label{four}

In order to provide a more detailed characterization of the macroscopic as well
as of the microscopic dynamics, in this section we analyse the Lyapunov spectra
for the fully-coupled network and for its disordered variants.

\subsection{Globally Coupled Network}

\begin{figure}[t!]
\includegraphics[draft=false,clip=true,height=0.5\textwidth]{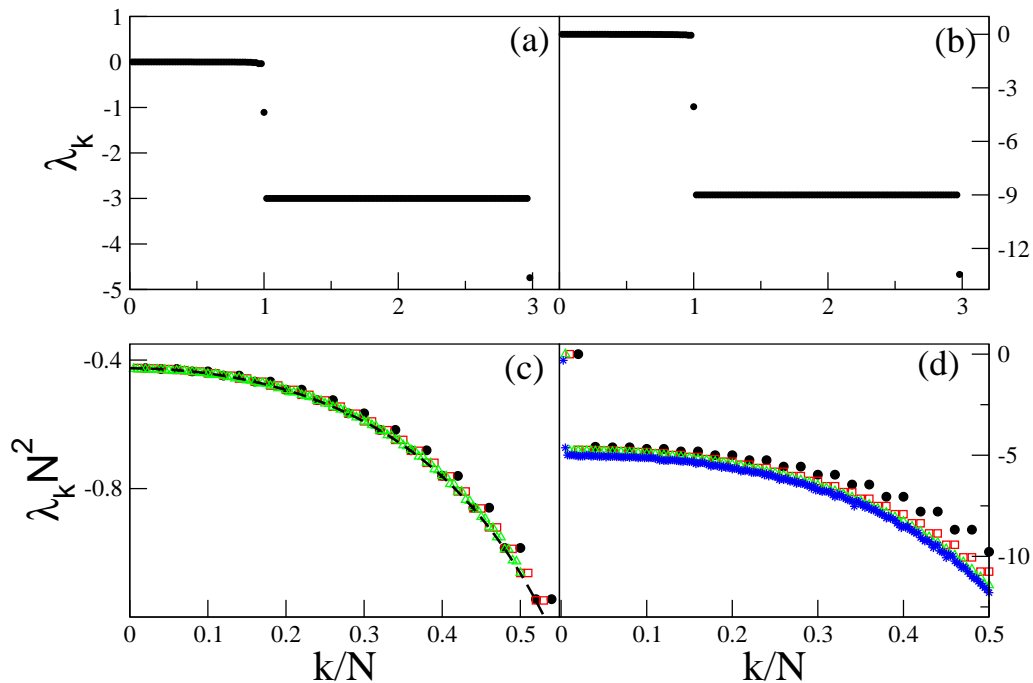}
\caption{
(Color online) Lyapunov exponents $\lambda_k$ versus $k/N$. Complete spectrum
for $N=50$ for $\alpha=3$ (a) and  $\alpha=9$ (b). Rescaled Lyapunov eigenvalues
of the first band $\lambda_k \times N^2$ for different  system sizes, namely $N=50$ (filled black
circles), 100 (open red squares), 200 (green open triangles) and 400 (blue
crosses), for $\alpha=3$ (c) and $\alpha=9$ (d). For $\alpha=3$ in (c) also the analytical
expression (dashed black line) reported in Eq. (29) in \cite{calamai} is shown.
The data has been obtained by following the evolution in the tangent space of the event driven map for  a
number of consecutive  spikes of the order of $10^8 -- 10^9$, after discarding a
transient composed by $50,000 \times N$ spikes.
}
\label{fig5}
\end{figure}

In this case, the fields seen by the neurons are equal to one another and
it is therefore sufficient to introduce a single pair of field variables $E$
and $Q$. The corresponding stability analysis for the splay state has been
analytically carried out in Ref.~\cite{zillmer2}, finding that the spectrum of
Floquet exponents is composed of a band of values of order $1/N^2$, which
become of order ${\mathcal O}(1)$ at one band extremum, plus two isolated
eigenvalues associated with the field dynamics. Therefore, it is convenient
to start the numerical analysis by testing the effect of attaching a pair of
field variables to each neuron. The results plotted in Fig.~\ref{fig5}a refer
to the Lyapunov spectrum of the splay state for $\alpha=3$ and $g=0.4$: one can
observe two bands and two isolated exponents. The first band, composed of $N-1$
nearly vanishing exponents, and the two isolated exponents coincide with the
eigenvalues analytically obtained in \cite{zillmer2}. The second band, composed
of $2(N-1)$ exponents quantifies the transversal stability of the
synchronization manifold $E_i=E_1$, $Q_i=Q_1$ $\quad \forall i$, where, without
loss of generality, we consider the field of the 1st neuron as the reference
variable. In fact, if one introduces the suitable difference variables,
\begin{equation}
w_{i}=E_{i}-E_{1} \qquad z_{i}=P_{i}-P_{1} \qquad i=2,\cdots,N \qquad .
\label{var_differenza}
\end{equation}
the corresponding tangent dynamics is ruled by the equations
\begin{eqnarray}\label{mappa_diff1}
\delta w_{i}(n+1) &=& \delta w_{i}(n)e^{-\alpha\tau(n)} +
   \delta z_{i}(n)\tau(n)e^{-\alpha\tau(n)} \\ \label{mappa_diff2}
\delta z_{i}(n+1) &=& \delta z_{i}(n)e^{-\alpha\tau(n)}  \qquad i=2,\cdots,N \quad .
\end{eqnarray}
This shows that the linearized dynamics of each pair of
twin variables $(\delta w_{i},\delta z_{i})$ is decoupled from the rest of the
network. Its stability can be evaluated by solving the corresponding
two-dimensional eigenvalue problem. Accordingly, we expect two bands of $N-1$
eigenvalues each. However, since the two eigenvalues of Eq.~(\ref{mappa_diff1})
are both equal to $-\alpha$, we do have a single band, as indeed observed in
Fig.~\ref{fig5}a. As a last check, we have verified the
$1/N^2$ scaling of the first band of the splay state. The nice overlap of the
three sets of exponents corresponding to $N=100$, 200, and 400 with 
the analytical estimate \cite{calamai} reported in
Fig.~\ref{fig5}c confirms the reliability of the numerical code.

The stability of CPOs arising when $\alpha>\alpha_c=8.34(1)$ \cite{vvres} is not
known. However, the above arguments concerning the presence of a single
degenerate band still hold, as confirmed by the spectrum plotted in
Fig.~\ref{fig5}b, which corresponds to $g=0.4$, and $\alpha=9$. This adds to
the first band that is again located just below zero, with the exception of the
first exponent which is exactly equal to zero, as a result of the
quasi-periodic dynamics of the single neurons (the second zero Lyapunov
exponent has been discarded while taking the Poincar\'e section).

Finally, in Fig.~\ref{fig5}d we have plotted the Lyapunov exponents of the first
band multiplied by $N^2$. The tendency to overlap of the spectra obtained for
$N=50,100,200,$ and 400 indicate that we are again in the presence of a
$1/N^2$ scaling as for the splay states, although finite-size corrections are
more relevant in this case. Accordingly, we conjecture that the $1/N^2$
dependence is a general property, which depends on the shape of the velocity
field (see \cite{calamai}), rather than on the structure of the solution itself.

\subsection{Diluted Network}

Since the collective solutions observed in the globally coupled network 
are characterized by many weakly stable directions, it is reasonable to expect
that generic perturbations of the network dynamics yield a chaotic
evolution. On the other hand, in the thermodynamic limit we expect the
inhomegenities induced by a random dilution to vanish.
In fact, we have already seen (in the previous section) that a macroscopically
periodic dynamics is eventually recovered. 
 
\begin{figure}[t!]
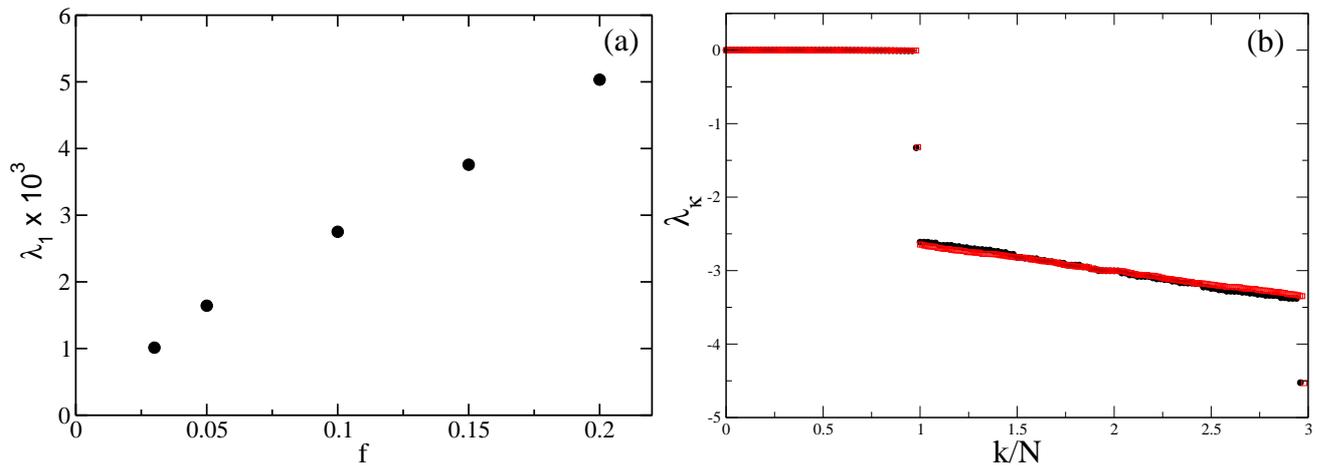

\includegraphics[draft=false,clip=true,height=0.34\textwidth]{FIG/fig6a}
\includegraphics[draft=false,clip=true,height=0.335\textwidth]
{FIG/fig6b}
\caption{
(a) Maximal Lyapunov exponent $\lambda_1$ as a function of the percentage $f$ of
broken links for $N=800$ and $\alpha=9$. (b) Lyapunov spectrum $\{\lambda_k\}$
versus $k/N$ for $\alpha=3$ with networks of size $N=50$ (filled black circles)
and $N=100$ (open red squares) and 20\%  of broken links.  The data has been
obtained by following the evolution of the event driven map and of the
associated linearized equations ruling the evolution  of the Lyapunov vectors
for $3-4 \times 10^8$ consecutive spikes, after discarding a transient composed
by $300,000 - 600,000$ spikes. The data refers to quenched disorder. }
\label{fig6}
\end{figure}

First of all we have verified that the network dynamics is chaotic for
finite $N$, as soon as $f>0$. In Fig.~\ref{fig6}a one can appreciate the growth
of the maximal (positive) Lyapunov exponent with the fraction of missing links
$f$ (in the quenched case).

For $\alpha=3$, analogously to the globally coupled case, the spectrum
is composed of two bands and two isolated eigenvalues (see Fig.~\ref{fig6}b).
On the one hand, the degeneracy among the exponents lying in the second band is
lifted (as a consequence of the disordered network structure) and the second
band acquires a finite width. On the other hand, the comparison between the
spectrum obtained for $N=50$ and $N=100$, reveals that the band width shrinks
with $N$ around the value $\lambda=-\alpha$. This result again confirms that
the inhomogenities of the disordered network vanish in the thermodynamic
limit. A similar scenario occurs also in the CPO regime, as well as for annealed
disorder.

\begin{figure}[t!]
\includegraphics[draft=false,clip=true,height=0.4\textwidth]{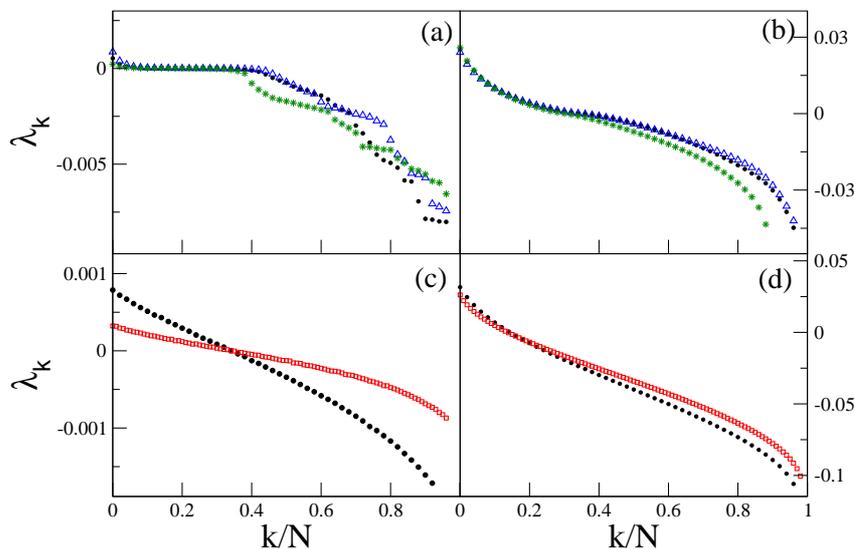}
\caption{
First band of the Lyapunov spectra $\{\lambda_k\}$ versus $k/N$ for $g=0.4$ 
with 20\% of broken links for quenched disorder with $\alpha=3$ (a) and 9 (b)
and for annealed disorder with $\alpha=3$ (c) and 9 (d). 
For the quenched disorder only spectra corresponding to system size $N=50$ are shown
but for three different random network configurations, while in the
annealed case spectra for network sizes $N=50$ (filled black circles) 
and $N=100$ (open red squares) are reported.
The data has been obtained by following the evolution of the
system and of its linearized copies for time lags similar to those reportd 
in the caption of Fig.~\ref{fig6}.
}
\label{fig7}
\end{figure}

Furthermore, in Fig.~\ref{fig7} (where we plot the first band of the
Lyapunov spectrum for quenched and annealed disorder, in both dynamical phases),
we see a variable number of positive exponents, indicating that finite
disordered networks are typically chaotic. More precisely, panels $a$ and $b$ of
Fig.~\ref{fig7}, refer to quenched disorder. In both cases, we present
the spectra resulting from three different realizations of the disorder.
We see that sample-to-sample fluctuations are quite relevant in the second
part of the band, while they affect less the largest exponents. Another
qualitative observation concerns the shape of the spectrum that seems to
be smoother in the presence of CPOs. Unfortunately, it is almost impossible to
perform a quantitative scaling analysis of the spectrum (given the need to
average over different realizations and to consider yet larger network sizes).
In the annealed case, Fig.~\ref{fig7}c/d, there is no need to average over
different realizations of the disorder and the Lyapunov spectra turn out to be
smooth. However a scaling analysis is still beyond our computer capabilities:
in both cases we report the spectra obtained for $N=50$ and $N=100$ which
are far from exhibiting a clean scaling behavior (for this reason we do not
even dare to formulate a conjecture).

\begin{figure}[ht]
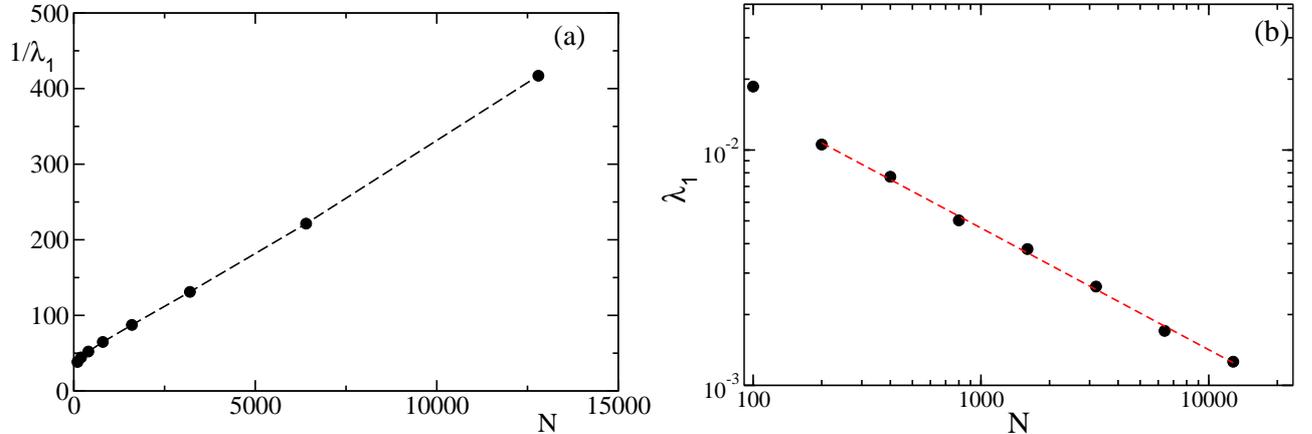

\includegraphics[draft=false,clip=true,height=0.32\textwidth]{FIG/fig8a}
\includegraphics[draft=false,clip=true,height=0.32\textwidth]{FIG/fig8b}
\caption{
Maximal Lyapunov exponent $\lambda_{1}$ versus $N$:
(a) annealed case; (b) quenched case. In the quenched
case also a power-law fit is reported (red dashed line)
with decay exponent $\beta = 0.51 \pm 0.01$.
The data has been obtained
by following the evolution in the tangent space of the event driven map for
$10^8 -- 10^9$ consecutive spikes, after discarding
a transient of $200,000$ spikes.
Both figures refer to 20 \% of dilution, with parameters
$\alpha= 9$, $g=0.4$ and $a=1.3$.
}
\label{fig8}
\end{figure}

Given the difficulties of dealing with the whole Lyapunov spectrum, we have
concentrated our efforts on the maximum exponent $\lambda_1$, since we are
thereby able to study larger networks. We limited ourselves to studying
the CPO regime. As shown in Fig.~\ref{fig8}, $\lambda_1$ decreases with $N$ as a
power law: in the annealed case, $\lambda_{1} \approx 1/N$, while in
the quenched case $\lambda_{1} \approx 1/\sqrt{N}$. This is at variance with
the Kuramoto model, where in the ``equivalent" quenched case, a $1/N$ behaviour
has been observed \cite{maistrenko}. Accordingly, although annealed networks
are more chaotic than the quenched ones over the numerically accessible network
sizes, we conclude that the opposite will be true for yet larger networks.

\section{Conclusions and perspectives}
\label{five}

Our numerical analysis suggests that in the thermodynamic limit, 
a random, uncorrelated network behaves like a homogeneous globally coupled
system with a rescaled value of the coupling constant to account for
the different fraction of active links. This is because each neuron receives
a random series of spikes: when the number of neurons increases, the number of
spikes per unit time increases as well, while the relative fluctuations
decrease and all neurons are affected by increasingly similar forcing fields 
(for a discussion on the thermodynamic limit in neural networks see also
\cite{hansel,vogels}).
While there are little doubts that the collective motion is independent of
the presence of disorder, less compelling is the evidence that the same is
true for the stability properties. This is because numerical simulations
are computationally expensive and it is not possible (at least for us)
to study the scaling behavior of the entire Lyapunov spectrum in disordered
systems. This is doable in homogeneous networks, thanks also to a faster
convergence to the thermodynamic limit, and we have indeed observed that
the first band of the Lyapunov spectrum scales as $1/N^2$, exactly as in the
splay state, a case that is analytically known \cite{calamai}. 
In disordered networks we have nevertheless been able to study the scaling
behaviour of the maximal Lyapunov exponent, finding that it is positive and
scales as $1/N^\beta$, with $\beta$ close to $1/2$ (resp. 1) in the quenched
(resp. annealed) case.
This means that deterministic chaos disappears in the thermodynamic limit.
This observation is consistent with the fact that the collective motion is
periodic: a periodically forced phase oscillator (such as a LIF neuron)
cannot be chaotic. However, nothing seems to prevent the collective motion
itself to be chaotic. Whether different connection topologies have to be
invoked or yet unidentified constraints forbid this to occur, remains as an
open question.

\acknowledgments
We acknowledge useful discussions with P. Bonifazi, M. Timme, and M. Wolfrum.
This work has been partly carried out with the support of the
EU project NEST-PATH-043309 and of the italian project ``Struttura e dinamica di reti
complesse'' N. 3001 within the CNR programme ``Ricerca spontanea a tema libero''. 


\end{document}